\documentclass{llncs}

\usepackage{etex} \reserveinserts{28}

\usepackage[english]{babel}
\usepackage{xcolor}
\usepackage{hyphenat}

\newcommand\isExercisedWithinAnalysis{is\-Exercised\-Within\-Analysis}

\usepackage{listings}
\usepackage{color}
\usepackage{latexsym}
\usepackage{mathtools}

\usepackage{amsthm}
\newcommand\xqed[1]{%
  \leavevmode\unskip\penalty9999 \hbox{}\nobreak\hfill
  \quad\hbox{#1}}
\newcommand\closeExample{\xqed{$\triangle$}}

\usepackage{amsfonts}
\usepackage{graphicx}
\usepackage{flushend}
\usepackage{siunitx}
\usepackage{bigfoot}
\usepackage{hyperref}

\usepackage[linesnumbered,noend]{algorithm2e}
\usepackage{cleveref}

\definecolor{dkgreen}{rgb}{0,0.6,0}
\definecolor{gray}{rgb}{0.5,0.5,0.5}
\definecolor{mauve}{rgb}{0.58,0,0.82}

\title{Verification Coverage}
\author{R. Casta\~no\inst{1}, V. Braberman\inst{1}, D. Garbervetsky\inst{1}, S. Uchitel\inst{1,2}\\
}

\institute{
Departamento de Computaci\'on, FCEN, Universidad de Buenos Aires, Argentina
    \and
    Department of Computing, Imperial College London, UK
    \email{\{rcastano,vbraber,diegog,suchitel\}@dc.uba.ar}
}

\newtheorem{mydef}{Definition}

\begin{document}

\maketitle
\begin{abstract}
    Software Model Checkers have shown outstanding performance improvements in recent times.
    Moreover, for specific use cases, formal verification techniques have shown to be highly effective, leading to a number of high\hyp profile success stories.
    However, widespread adoption remains unlikely in the short term and one of the remaining obstacles in that direction is the vast number of instances which software model checkers cannot fully analyze within reasonable memory and CPU bounds.
    The majority of verification tools fail to provide a measure of progress or any intermediate verification result when such situations occur.

    Inspired in the success that coverage metrics have achieved in industry, we propose to adapt the definition of coverage to the context of verification.
    We discuss some of the challenges in pinning down a definition that resembles the deeply rooted semantics of test coverage.
    Subsequently we propose a definition for a broad family of verification techniques: those based on Abstract Reachability Trees.
    Moreover, we discuss a general approach to computing an under\hyp approximation of such metric and a specific heuristic to improve the performance.

    Finally, we conduct an empirical evaluation to assess the viability of our approach.
\end{abstract}

\hyphenation{key-span CPA-check-er is-Exercised-Within-Analysis}

\section{Introduction}

Test coverage in its different flavors is widely adopted and discussed~\cite{whittaker2012google,page2008we,motor2008misra} by major companies with regards to development and testing methodologies.
Other software verification tools and techniques~\cite{newcombe2015amazon,ball2011decade} have also been successfully used in industry for either full verification or bug finding.

However, verification tools sometimes produce inconclusive results, that is, they neither find errors nor completely prove the absence of a specific bug within the allotted time or resource bounds.
In these cases, most tools fail to provide any safety assurances, making the time spent on verification entirely useless.

In this work, we adapt the definition of code coverage used in testing to a broad family of verification algorithms.
Our definition attempts to replicate the semantics of code coverage in testing, so as to extend the deep-rooted understanding of the concept to the context of verification.

The main contribution of this work is a novel perspective on quantifying progress and safety assurances of incomplete verification attempts.
We discuss a number of approaches to producing an under\hyp approximation of the coverage metric we put forth and also propose a heuristic to improve the efficiency.
We empirically evaluate a proof\hyp of\hyp concept implementation on benchmark instances and discuss the results.
Finally, we end the discussion with an overview of related works and a few concluding remarks.

\section{Motivation}
\label{sec:motivation}

    \subsection{Why quantify incomplete verification coverage?}

    Verification encompasses a wide range of techniques and the reasons for producing an inconclusive result are equally as varied.
    Some approaches, such as Corral~\cite{lal2012solver} and many other flavors of BMC~\cite{biere1999symbolic}, selectively explore a portion of the possible behaviors of a system.
    In many cases, BMC-based tools can normally terminate the execution without a conclusive result.

    Many others attempt to exhaustively explore a possibly infinite state space.
    This can lead to long running explorations which, in turn, could case exhaustion of system resources, including memory and CPU time budget.

    Moreover, many tools can fail throughout verification even after performing significant work.
    For example, some techniques rely on finding an interpolant to a formula in order to refine an abstraction.
    But there are fundamental and practical limitations to the underlying logic engines that produce the interpolants, therefore, this process can fail.
    In this setting, as long as the solver is capable of generating interpolants, abstraction refinement continues, possibly for many iterations, and then suddenly stops without a conclusive result.

    In all these cases, despite the inconclusive results, the verification attempt analyzed a number of system behaviors.
    Similar to a test suite, an incomplete verification attempt can help increase the confidence in the correctness of the system and an adequate measure could be used as part of a dependability case.

    \subsection{Challenges in defining verification coverage}

   Test suite adequacy is frequently described with respect to code coverage.
   However, the usual metrics used in testing are not immediately applicable to verification.

   Code coverage in testing is defined in terms of the parts of the system exercised throughout the execution of at least one test case.
More specifically, statement coverage is defined as~\cite{young2005software}:
\begin{quote}
	$C_{Statement}$ of $T$ for $P$ is the fraction of statements of program P executed by at least one test case in $T$.

	\begin{equation*}
	C_{Statement} = \frac{\text{number of executed statements}}{\text{number of statements}}
	\end{equation*}
\end{quote}

One of the difficulties of adapting this definition is that, in the context of testing, test cases are expected to run to completion but that notion cannot always be directly translated to partial verification attempts.
More precisely, verification techniques frequently employ abstractions that do not necessarily capture any execution in its entirety.
This can be misleading, because no relevant assertions might have been reached by the time the verification attempt was interrupted and assertion failures related to the explored statements, even if trivial, would not have been caught.

\begin{example}[Loop unrolling]
\begin{figure}
\begin{lstlisting}
int nondet();
#define false 0;
int main() {
  for (i = 0; i < 1000000; i++);
  assert(false);
  return 0;
}
\end{lstlisting}
\caption{Long running loop.}
\label{fig:non_linear}
\end{figure}

Let's consider the example in Figure~\ref{fig:non_linear} and an exploration using value analysis, a verification technique that explores paths in the Control Flow Automaton (CFA) of the program while keeping track of the actual values of variables, i.e. value analysis.
    
    The \emph{for}\hyp loop runs 1000000 times increasing the value of \emph{i} to that number.
    In this case, with a short enough time limit, the algorithm would keep unwinding the loop until timing out.
    This situation leads to an explored state space that contains no complete executions of the code.
    The analysis would be oblivious to the assertion in any incomplete verification attempt that does not fully unwind the loop.

    Conversely, if we considered the exercised statements as covered, a relatively high coverage would be misleading, when the trivially \emph{false} assertion would be reached by the only possible execution of the code.
\closeExample
\end{example}

When defining a notion of verification coverage, one might be tempted to quantify progress by directly measuring the internal representation of the technique as it is.
In fact, this is the implementation of the coverage measure reported by CPAchecker\cite{beyer2011cpachecker}.
However, the abstractions commonly used can capture spurious behaviors due to over\hyp approximation.
Consequently, it could be misleading to quantify progress this way.

\begin{example}[Unreachable code]
\begin{figure}
\begin{lstlisting}
int nondet();
int main() {
    int x = nondet();
    int reached_dead_code = 0;
    int z = 1;
    if (x*x < 0) {
        reached_dead_code = 1;
    } else {
        z = 1;
    }
    %*\label{line:assertion}*)assert(!reached_dead_code);
    return 0;
}
\end{lstlisting}
\caption{Dead code.}
\label{fig:dead_code}
\end{figure}

The code snippet in Figure~\ref{fig:dead_code} contains unreachable code.
The condition in the \emph{if} statement never holds, that is, the \emph{then} branch can never be executed, therefore the assertion in line~\ref{line:assertion} always holds.

In this case, a verification attempt using lazy predicate analysis would initially deal with an abstract representation which would not be precise enough to prove the \emph{then} branch unreachable.
This first attempt would find a spurious counterexample and attempt to refine the abstraction.
However, abstraction refinement, i.e. finding the right predicate to prove the \emph{then} branch unreachable, would fail due to the non\hyp linear condition.

In this case, the unreachable portion of the state space corresponds to dead code, but more subtle cases can, and often do, arise in practice.
Therefore, unless unreachable states are ignored, the coverage metric can be arbitrarily inflated with respect to its intuitive interpretation.
\closeExample
\end{example}

%

\section{Coverage definition for Abstract Reachability Trees}

In this section, we will present a definition of coverage for Abstract Reachability Tree~\cite{henzinger2002lazy} (ART)\hyp based techniques.
In order to make the presentation self\hyp contained, we will first provide the necessary background.

\subsection{Background}
ARTs are a widely used data structure in verification.
A number of dissimilar techniques have been implemented using ARTs, including lazy predicate abstraction~\cite{henzinger2002lazy}, BMC~\cite{biere1999symbolic}, value analysis~\cite{beyer2013explicit}, IC3\hyp based techniques~\cite{cimatti2012software} and CEGAR variants of the former, among other.
As such, targeting ARTs allows us to handle a wide variety of techniques, at least in their ART implementation.

Moreover, Conditional Model Checking~\cite{beyer2011conditional,beyer2012conditional} (CMC) has been instantiated for ART\hyp based techniques and implemented within CPAchecker~\cite{beyer2011cpachecker}, providing valuable infrastructure.
CMC proposes to augment software model checkers by either returning a counterexample to the property of interest or returning a condition $\psi$ under which the model checking attempt proved the program safe to execute.

The instantiation of CMC for ART\hyp based techniques represents $\psi$ in terms of Assumption Automata over the alphabet of statements.
That is, $\psi$ in the case of ARTs is a predicate over sequences of statements. 

An Assumption Automaton captures the structure of the ART at the moment when verification was interrupted: with the exception of a distinguished state, each state in the Assumption Automaton corresponds to a single state in the Control Flow Automaton of the system\hyp under\hyp verification and transitions are labeled with statements.
Lastly, the distinguished state \texttt{FALSE} captures the unexplored state space.

Consequently, $\psi(\pi)$ holds \emph{iff} $\pi$ does not constitute a path from the initial state to the distinguished state \texttt{FALSE} in the Assumption Automaton.

\subsection{Definition}

Our definition aims to tackle the issues discussed and exemplified in Section~\ref{sec:motivation} by adapting the notion of running a test to completion to the context of verification.

\begin{mydef}
    \label{def:coverage}
	Given a predicate $\psi$ that corresponds to the output of a conditional model checker and a set $\mathcal{T}$ of terminating executions of the system\hyp under \hyp verification, i.e. feasible paths from an initial state in the CFA and reaching the system exit, a statement $s$ is considered to be covered by the exploration when the following holds:

	$\exists \; t \in \mathcal{T}$, such that $\varphi(t) \land \isExercisedWithinAnalysis_{\psi}(s, t)$

	Moreover, in the context of ART-based analyses, the predicate $\isExercisedWithinAnalysis_{\psi}(s,t)$ holds iff:

	\begin{equation*}
		\exists \; \pi \in \Sigma^*\text{, such that }isPrefix(\pi, t) \land s \in \pi \land \psi(\pi)
	\end{equation*}

	where $isPrefix(\pi, t) = \exists \pi' \in \Sigma^* \text{ such that } \pi \cdot \pi' = t$ and $\cdot$ stands for concatenation.

\end{mydef}

Intuitively, \Cref{def:coverage} states that a statement $s$ is considered covered when some terminating execution ($t$) satisfying the safety property ($\varphi(t)$) contains an analyzed ($\psi(p)$) prefix ($isPrefix(p, t)$) that exercises $s$ ($s \in p$).

Having adapted the notion of a particular statement being covered in the context of verification, statement coverage in verification can then be defined as:

$C_{Statement}(\psi)$ for $P$ is the fraction of statements of program P covered by the exploration captured by $\psi$.

	\begin{equation*}
        C_{Statement}(\psi) = \frac{\text{number of statements covered by }\psi}{\text{number of statements}}
	\end{equation*}

\section{Computing coverage}
\label{sec:computing_coverage}
\begin{algorithm}[!ht]
\SetKwInOut{Input}{input}
\SetKwInOut{Output}{output}
\Input{
    An Assumption Automaton: \texttt{AA},
    A system \texttt{S}
}
 \Output{A number indicating statement coverage.}
%
 \texttt{allStatements} := statements(\texttt{S})

 \texttt{covered} := $\emptyset$

 \texttt{remaining} := \texttt{allStatements}
 
 \While{\texttt{remaining} $ \neq \emptyset$}{\label{alg:loop_head}
     \texttt{specification} := \emph{terminatingExecutionSpecCovering}(\texttt{remaining}, \texttt{AA})

     \texttt{result} := \emph{verify}(\texttt{S}, \texttt{specification})

     \uIf{\texttt{result.unknown}}{
        warning(``Producing under-approximation.'')\label{alg:unknown}

        break
     }
     \uElseIf{$\neg$ \texttt{result.foundCounterexamples}}{
        break\label{alg:none_found}
     }
     \Else{
         \texttt{cex} := \texttt{result.counterexamples}

         \texttt{covered} := \texttt{covered} $\cup$ \emph{exercisedWithinAnalysis}(\texttt{cex}, \texttt{AA})

         \texttt{remaining} := \texttt{remaining} $\setminus$ \texttt{covered}
     }
 }
 \Return{\texttt{covered}/\texttt{allStatements}}
 \caption{Computing Coverage}
\label{alg:exact_coverage}
\end{algorithm}

We propose to compute verification coverage by encoding, as a safety property, the negation of the conditions under which a statement is covered.
By feeding the safety property to an off\hyp the\hyp shelf verification tool, the counterexamples produced confirm that the statement is covered whereas, if the output indicates the property holds, then the statement is not covered.
Concretely, our algorithm takes as input an Assumption Automaton and outputs the verification coverage.

As shown in Figure~\ref{alg:exact_coverage}, we iteratively augment the set \texttt{covered} until either the verification task yields no counterexample (line~\ref{alg:none_found}) or all statements have been covered (loop head, line~\ref{alg:loop_head}).
When the former occurs, statements in \texttt{remaining}, the complement of \texttt{covered} are not covered by the exploration captured in the Assumption Automaton.

Our algorithm yields the exact statement coverage as long as the procedure \emph{verify} always return a conclusive result.
In practice, this is not always the case, as anticipated in line~\ref{alg:unknown}, therefore many times the algorithm produces an under\hyp approximation.
It is worth noting that if an exact result is not desired, it is possible to add a time budget or a limit in the number of iterations of the main loop and the result is guaranteed to be an under\hyp approximation.
We will discuss a generic approach to producing under\hyp approximations in Section~\ref{sec:under_approx}

The specifics of how to build the corresponding specification depend on the desired verification technique and chosen tool.

Lastly, \emph{exercisedWithinAnalysis} instantiates the predicate \emph{\isExercisedWithinAnalysis}, as stated in \Cref{def:coverage}.

The function \emph{exercisedWithinAnalysis} is implemented by interpreting the counterexample as a path in the Assumption Automaton.
The automaton contains a distinguished state \texttt{FALSE}, which captures the unexplored state space, and given the counterexample, i.e. a sequence of statements capturing a system execution, it is possible to follow, from an initial state, the transitions corresponding to each statement of the execution.
The statements corresponding to transitions before reaching \texttt{FALSE} will be included in the result of the procedure.

\begin{mydef}
    Given a sequence of statement \texttt{cex}, the following holds for the result of \emph{exercisedWithinAnalysis}:
    \begin{align*}
    s \in exercisedWithinAnalysis(\texttt{cex}, \texttt{AA}) \; \emph{iff} \\ 
    \isExercisedWithinAnalysis_{\psi}(s, \texttt{cex})
    \end{align*}
\end{mydef}

\section{Under\hyp approximating coverage}
\label{sec:under_approx}

Algorithm~\ref{alg:exact_coverage} generates the exact coverage under ideal conditions, but might incur in a significant performance penalty in doing so.
Consequently, in many cases it is impractical to compute the exact coverage  and an under\hyp approximation can provide value.

Our approach to computing a verification coverage under\hyp approximation consists of generating a number of executions and then, for each of them, computing which of the statements exercised remain within the explored state space.

The input to our algorithm is an Assumption Automaton representing the progress achieved by an interrupted verification attempt and our output consists of an under\hyp approximation of the verification coverage.

Calculating the number of statements within the explored region is fairly straightforward, as explained in Section~\ref{sec:computing_coverage}: this is implemented in procedure \emph{exercisedWithinAnalysis} for Algorithm~\ref{alg:exact_coverage}.

On the other hand, even though a number of suitable existing techniques can generate system executions, it might be necessary to tweak them for the specific use case of under\hyp approximating a metric of coverage.

In principle, it would be possible to leverage random testing or logging of the running system to fulfill the goal of gathering system executions, but we will only focus on a verification\hyp based approach.

As explained in the previous section, we can use standard verification techniques to generate system executions.
To do so, we can create a specification asserting that no executions reach the program exit.
A feasible counterexample would then constitute a terminating execution, as intended.

One caveat with all these standard approaches is that they fundamentally ignore the end goal, which is to compute a good under\hyp approximation of the verification coverage metric. 

\subsubsection{Custom trace prioritization in ART-based verification}

\hfill~\linebreak
The idea of using standard verification to produce system executions is appealing to us due to its simplicity and ease of implementation.

However, there seems to be room for improvement: the information captured by the Assumption Automaton of the incomplete verification attempt is not leveraged to maximize the coverage under\hyp approximation.

We implemented a heuristic that aims to prioritize the generation of traces that maximize the coverage measure.
Our heuristic is compatible with any ART\hyp based analysis.

The heuristic consists in assigning a tentative coverage score to each state in the Assumption Automaton and using the score to guide the exploration of ART nodes.
In order to compute that score we compose the Assumption Automaton with the CFA and calculate the least fixed point of the function $Reach ((s_a, s_{CFA}))$, where $s_a$ and $s_{CFA}$ states of the Assumption Automaton and the CFA, respectively and $(s_a, s_{CFA})$ are the states of the composition:

\begin{align*}
    Reach((s_a, s_{CFA})) = 
    \begin{cases}
        s_{CFA} \cup \bigcup\limits_{(s'_a, s'_{CFA}) \in succ((s_a, s_{CFA}))} Reach((s'_a, s'_{CFA})) & \text{ if } s_a \neq \texttt{FALSE} \\
        \emptyset & \text{otherwise}
    \end{cases}
\end{align*}

The size of the largest $Reach((s_a, s_{CFA}))$ will constitute the tentative score assigned to state $s_a$.

Intuitively, the least fixed point will tend to propagate backwards the lines to be potentially covered.
Accordingly, if ART construction is prioritized taking into account this score, states that potentially cover more lines will tend to be expanded first.

Concretely, we will use the tentative coverage score to tweak standard traversals such that the exploration is slightly guided toward more promising regions of the state space. 

%

\section{Evaluation}

We conducted a round of initial experiments with the exact algorithm, but the time necessary to compute the coverage measure was, in many cases, a multiple of the original verification time.
This prompted us to develop the under\hyp approximation algorithm discussed.
In this section we discuss the preliminary empirical evaluation we performed.

Our main research questions are the following:

\begin{itemize}
    \item {\bf RQ 1}: Is it possible to compute a useful under\hyp approximation of verification coverage efficiently?
    \item {\bf RQ 2}: Does the heuristic proposed improve the under\hyp approximation produced within a limited time budget?
\end{itemize}

Our experiments were composed of two phases.
The first phase consisted of running a verification algorithm with a time limit of \SI{900}{s}, which is standard in verification competitions~\cite{DBLP:conf/tacas/Beyer16}.
For this phase, we used 2 different techniques, lazy predicate abstraction and value analysis.
We try unrelated technique in the first phase to evaluate how our approach performs in different settings.

The instances we used are those previously chosen to evaluate CMC, belonging to the families SystemC and DeviceDrivers of the SV-COMP set of benchmark instances.

The second phase consisted in computing, for the instances that did not produce a conclusive result, an under\hyp approximation of the statement coverage using verification to generate complete executions.
We used two versions of the same technique for the second phase: a standard verification technique, which serves as a baseline, and a slight modification that integrates our heuristic.
The technique that we used is a flavor of ART\hyp based value analysis with a traversal geared towards quickly reaching the end of the program.
The traversal is a combination of depth\hyp first\hyp search with postorder (CFA nodes with a lower postorder index are selected first).

We integrate our heuristic by using it as a second criterion for selection in the traversal, that is, postorder is considered first and only in case of a tie the score produced by our heuristic is used.

We configured our implementations with the lower of two limits: a time limit of \SI{900}{s} and a limit of 10 counterexamples.

\Cref{table:explicit,table:predicate} contain the results of the second phase of the experiments.
Both tables comprise the instances that failed to produce a conclusive result within \SI{900}{s}: in the case of \Cref{table:explicit}, the verification attempt used value analysis, whereas \Cref{table:predicate} corresponds to verification using lazy predicate abstraction.
Both techniques can be configured in a number of different ways, we used the same configuration used for the evaluation of CMC\footnote{CMC evaluation is available online: \url{https://www.sosy-lab.org/~dbeyer/cpa-cmc/}}.

The columns hold the following information, values enclosed in parenthesis correspond to the heuristic presented:
\begin{itemize}
\item \# lines: Total number of lines excluding comments and blank lines.
\item Over A.: Over\hyp approximation of coverage reported by CPAchecker.
    Corresponds to the number of lines for which a node existed in the ART at the moment when the execution was interrupted.
    The ART nodes could be unreachable, therefore this constitutes an over\hyp approximation of our measure of coverage.
\item Under A.: Statement coverage under\hyp approximation we computed.
\item \# exec.: Number of executions generated, and subsequently used, to compute the under\hyp approximation.
\item Bug?: This column contains a check mark ($\checkmark$) when, while attempting to find a terminating execution, the verification tool instead found and confirmed an assertion violation.
\item CPU time: Total CPU time used.
    We use a soft limit of \SI{900}{s} but allow at most 100 additional seconds for the CPAchecker to return.
\end{itemize}

\begin{table*}
	\centering
	\begin{tabular}{l|c|c|c|c|c|c|l}
    &      &   & \multicolumn{4}{c|}{Baseline ( with {\bf Heuristic})} \\ 
 & \# lines & Over A. & Under A. & \# exec. & Bug? & CPU time & \\
token\_ring.14.BUG.c 	& 833 	& 594 	& 570 ({\bf -}) 	& 1 ({\bf -}) 	& - ({\bf -}) 	& 28.68s ({\bf 906.10s}) 	&  \\
toy.c 	& - 	& - 	& - ({\bf -}) 	& - ({\bf -}) 	& - ({\bf -}) 	& 979.53s ({\bf 971.91s}) 	&  \\
transmitter.16.BUG.c 	& 900 	& 658 	& - (N/A) 	& - ({\bf 1}) 	& - ({\bf \checkmark}) 	& 921.54s ({\bf 22.85s}) 	&  \\
farsync.BUG.c 	& 5700 	& 1944 	& - (N/A) 	& - ({\bf 7}) 	& - ({\bf \checkmark}) 	& 905.54s ({\bf 217.85s}) 	&  \\
gigaset.BUG.c 	& 16188 	& 1609 	& 1077 ({\bf 970}) 	& 9 ({\bf 9}) 	& - ({\bf -}) 	& 55.91s ({\bf 902.60s}) 	&  \\
lirc\_imon.BUG.c 	& 2166 	& 875 	& - ({\bf 611}) 	& - ({\bf 6}) 	& - ({\bf -}) 	& 907.52s ({\bf 905.77s}) 	&  \\
loop.BUG.c 	& - 	& - 	& - ({\bf -}) 	& - ({\bf -}) 	& - ({\bf -}) 	& 902.31s ({\bf 912.45s}) 	&  \\
ppp\_generic.BUG.c 	& 6969 	& 1849 	& - ({\bf 848}) 	& - ({\bf 2}) 	& - ({\bf -}) 	& 923.36s ({\bf 924.99s}) 	&  \\
synclink\_gt.BUG.c 	& 11780 	& 1112 	& - (N/A) 	& - ({\bf 1}) 	& - ({\bf \checkmark}) 	& 910.66s ({\bf 13.68s}) 	&  \\
 &     &   & & & & \\

\end{tabular}

	\caption{Complete results from explicit AA}
    \label{table:explicit}
\end{table*}

\begin{table*}
	\centering
	\begin{tabular}{l|c|c|c|c|c|c|l}
    &      &   & \multicolumn{4}{c|}{Baseline (with {\bf Heuristic})} \\ 
 & \# lines & Over A. & Under A. & \# exec. & Bug? & CPU time & \\
kundu.c 	& 266 	& 256 	& 250 ({\bf 252}) 	& 1 ({\bf 2}) 	& - ({\bf -}) 	& 15.16s ({\bf 878.60s}) 	&  \\
mem\_slave\_tlm.3.c 	& 796 	& 550 	& 490 ({\bf 490}) 	& 1 ({\bf 2}) 	& - ({\bf -}) 	& 20.46s ({\bf 856.79s}) 	&  \\
mem\_slave\_tlm.4.c 	& 801 	& 553 	& 495 ({\bf 495}) 	& 1 ({\bf 2}) 	& - ({\bf -}) 	& 26.74s ({\bf 856.69s}) 	&  \\
mem\_slave\_tlm.5.c 	& 806 	& 753 	& 500 ({\bf 500}) 	& 1 ({\bf 2}) 	& - ({\bf -}) 	& 32.46s ({\bf 864.80s}) 	&  \\
pipeline.c 	& 388 	& 267 	& 0 ({\bf 0}) 	& 0 ({\bf 0}) 	& - ({\bf -}) 	& 87.67s ({\bf 867.81s}) 	&  \\
token\_ring.03.c 	& 306 	& 261 	& 227 ({\bf 253}) 	& 1 ({\bf 5}) 	& - ({\bf -}) 	& 17.94s ({\bf 854.81s}) 	&  \\
token\_ring.04.c 	& 364 	& 307 	& 265 ({\bf 303}) 	& 1 ({\bf 6}) 	& - ({\bf -}) 	& 60.42s ({\bf 861.16s}) 	&  \\
token\_ring.05.c 	& 422 	& 361 	& 303 ({\bf 353}) 	& 1 ({\bf 7}) 	& - ({\bf -}) 	& 297.92s ({\bf 863.45s}) 	&  \\
token\_ring.06.c 	& 480 	& 415 	& 341 ({\bf 403}) 	& 1 ({\bf 8}) 	& - ({\bf -}) 	& 439.57s ({\bf 876.84s}) 	&  \\
token\_ring.07.c 	& 538 	& 469 	& - ({\bf 443}) 	& - ({\bf 8}) 	& - ({\bf -}) 	& 905.58s ({\bf 918.29s}) 	&  \\
token\_ring.08.c 	& 596 	& 523 	& - ({\bf 421}) 	& - ({\bf 2}) 	& - ({\bf -}) 	& 909.00s ({\bf 906.12s}) 	&  \\
token\_ring.09.BUG.c 	& 660 	& 582 	& 456 ({\bf 458}) 	& 1 ({\bf 1}) 	& - ({\bf -}) 	& 54.39s ({\bf 913.39s}) 	&  \\
token\_ring.14.BUG.c 	& 833 	& 743 	& 570 ({\bf -}) 	& 1 ({\bf -}) 	& - ({\bf -}) 	& 190.27s ({\bf 910.16s}) 	&  \\
toy.c 	& - 	& - 	& - ({\bf -}) 	& - ({\bf -}) 	& - ({\bf -}) 	& 967.02s ({\bf 973.23s}) 	&  \\
toy1\_BUG.c 	& 316 	& 295 	& N/A (N/A) 	& 10 ({\bf 1}) 	& \checkmark ({\bf \checkmark}) 	& 34.86s ({\bf 14.19s}) 	&  \\
pktcdvd.BUG.c 	& 6925 	& 4660 	& - ({\bf 857}) 	& - ({\bf 3}) 	& - ({\bf -}) 	& 903.28s ({\bf 855.53s}) 	&  \\
 &     &   & & & & \\

\end{tabular}

	\caption{Complete results from predicate AA}
    \label{table:predicate}
\end{table*}

\subsection{Assertion failures}

As a side effect of our definition requiring terminating executions, it is possible to find executions violating the property of interest when collecting execution for an under\hyp approximation.

This occurred in our experiments: for some of the instances, while attempting to produce terminating executions, the software model checker found assertion violations instead, as reported in column ``Bug?'' in \Cref{table:predicate,table:explicit}.

This is not central to our contribution, but is relevant to the interpretation of the results, since our under\hyp approximation algorithm stops when an assertion failure is reached.

\subsection{RQ1: Is it possible to compute a useful under\hyp approximation of verification coverage efficiently?}
As shown in \Cref{table:predicate}, our approach in conjunction with the proposed heuristic generated an under\hyp approximation for 14 out of 16 incomplete lazy predicate analysis explorations.
\Cref{table:explicit} shows similar results, producing an under\hyp approximation for 6 out of 9 instances.

We used a time limit of \SI{900}{s} which corresponds to the time allotted for the first phase, the one consisting of the actual verification attempt.
We consider that longer running times could be deemed impractical by potential users but, for most instances, an under\hyp approximation was produced before the time limit and the algorithm only kept refining it.

In this context, under\hyp approximations are useful when they approach the actual statement coverage value.
Only for three instances the algorithm produced a trivial under\hyp approximation of 0: in two cases (\texttt{transmitter.16.BUG.c} and \texttt{synclink gt.BUG.c}), it found an assertion failure and stopped within 5\% of the allotted time.
In the remaining case (\texttt{pipeline.c}) the program exit seems to be unreachable, since the instance resulting from replacing all existing assertions by a single trivially false assertion at the return of the main method can be proved safe using CPAchecker.

The values shown in \Cref{table:explicit} are mostly (9 out of 14) within 20\% below the over\hyp approximation, possibly even closer to the actual value.

\Cref{table:predicate} contains more varied results.
Nonetheless, all the values generated are within 50\% less than the over\hyp approximation produced by CPAchecker, if we consider the instances for which we produce an under\hyp approximation and exclude those for which an assertion was found.
These lower ratios between the under\hyp approximation and the over\hyp approximation are consistent with the inner workings of the technique.
Lazy predicate abstraction starts using a coarse predicate abstraction that keeps track of no predicates.
Throughout the construction of the ART, especially with a breadth\hyp first\hyp search traversal, a significant number of unreachable states could be added, inflating the over\hyp approximation calculated by CPAchecker.

In our opinion, these results suggest an affirmative answer to {\bf RQ1}: it seems possible to generate these under\hyp approximations efficiently since we were able to produce a result for the majority of the instances evaluated and the under\hyp approximations are of acceptable quality.

\subsection{ RQ 2: Does the heuristic proposed improve the under\hyp approximation produced within a limited time budget?}

The heuristic proposed produced an under\hyp approximation at least as good as the standard verification technique for 22 out of the 25 instances considered within the same allotted running time.
Moreover, the heuristic outperformed the baseline in 11 out of 25 instances.

These results seem to support an affirmative answer to {\bf RQ2} as well.

\section{Related Work}

There is a broad spectrum of research both proposing different test adequacy criteria and evaluating their relation with reliability~\cite{zhu1997software}.
However, to the best of our knowledge, the mentions of code coverage in the context of verification are far fewer.

The most related previous work defined a measure of verification coverage~\cite{arlt2014gradual} consisting of the number of statements proven safe at a certain point.
The definition has some subtleties, but a statement which cannot throw any of the predefined exceptions of interest is considered safe.
The semantics, in this case, is fundamentally different than that of test suite coverage making the metrics incomparable, but might also not be entirely intuitive or easy to interpret, as the authors themselves acknowledge, since many statements can be proved safe syntactically, regardless of the context.
In order to compute verification coverage, they propose to adapt a standard imprecise verification algorithm such that it computes a coverage metric besides producing the warnings of potential errors.
Their empirical evaluation suggests that the approach incurs in a significant performance penalty: the running time of the algorithm increases by an order of magnitude when adapted in this way.
Lastly, it is not straightforward how to compute the coverage measure for an incomplete verification attempt using a standard technique, as we intend in our setting.

Existing work discussing functional coverage~\cite{piziali2007functional} briefly mentions two possible high-level approaches to quantify the verification coverage of \emph{complete} verification attempts of a portion of a system.
Our work, in contrast, discusses the topic at a much lower level, having a single \emph{incomplete} verification attempt as our object of study.

Loop coverage has been discussed previously~\cite{lal2014powering}, although tangentially, in relation to bounded-exploration tools.
These tools can miss trivial defects when the bound prevents the exploration from crossing a long-running loop.
Loop coverage is not defined in that work and only its intuitive meaning is used.
The authors use defect recall as a proxy for loop coverage.
In contrast, we put forth a detailed definition and discuss at length different ways compute an under-approximation.

CPAchecker reports a coverage metric which, to the best of our knowledge, has not been published nor defined formally.
The number reported, as currently implemented, can report unreachable code as covered and corresponds to an over\hyp approximation of our coverage metric.
We discussed the problems of this semantics in Section~\ref{sec:motivation}.

Our work is heavily influenced by earlier work on partial verification attempts, such as the notion of Conditional Model Checking~\cite{beyer2012conditional}, exploration oracles~\cite{castano2016model} and other complementary representations~\cite{christakis2012collaborative}.
Their approaches, however, neither aim to quantify coverage nor provide any alternative metric of progress or exploration.

\section{Conclusion and Future Work}

Software model checkers have dramatically improved within the last few years and state\hyp of\hyp the\hyp art tools are capable of tackling an increasing number of industrial instances.
However, a large number of systems remain intractable for full verification.
In this work, we attempt to extract valuable information from incomplete verification attempts by producing a verification coverage metric.
We defined verification coverage in a manner that is consistent with that of coverage in the context of testing and discussed several ways of producing an under\hyp approximation of that metric.
Our empirical evaluation suggests that acceptable under\hyp approximations can be computed efficiently.

Future work includes analyzing the correlation between our definition of coverage and reliability, defect coverage and other relevant aspects of the quality of the system.

Moreover, it would be interesting to explore how verification coverage evolves throughout the execution of a software model checker.

It would also be valuable to propose and examine other quantifications of verification progress and to devise efficient ways to compute them.

\bibliographystyle{abbrv}
\bibliography{related}
\end{document}